\documentclass{aa}
\usepackage{graphicx}
\usepackage{txfonts}
\newcommand{\beq}{\begin{equation}}
\newcommand{\eeq}{\end{equation}}
\newcommand{\beqn}{\begin{eqnarray}}
\newcommand{\eeqn}{\end{eqnarray}}

\begin{document}
   \title{Quasi-linear diffusion driving the synchrotron emission
   in active galactic nuclei}

   \author{Z. Osmanov\inst{}\thanks{E-mail:
z.osmanov@astro-ge.org (ZO); g.machabeli@astro-ge.org (GM) } and G.
Machabeli$^{}$
          }

   \offprints{Z. Osmanov}

   \institute{Center for Theoretical Astrophysics, ITP, Ilia State University,
              Kazbegi str. $2^a$, Tbilisi, 0160, Georgia
             }



  \abstract
  {}
   {We study the role of the quasi-linear diffusion (QLD) in producing
   $X$-ray emission by means of ultra-relativistic electrons in AGN magnetospheric
   flows.}
   {We examined two regions: (a) an area close to the black hole
   and (b) the outer magnetosphere. The synchrotron emission has
   been studied for ultra-relativistic electrons and was shown that
   the QLD generates the soft and hard $X$-rays, close to the
   black hole and on the light cylinder scales respectively. }
   {By considering the cyclotron instability, we show that despite the short
   synchrotron cooling timescales, the cyclotron modes
   excite transverse and longitudinal-transversal waves. On the
   other hand, it is demonstrated that the synchrotron reaction force and
   a force responsible for the conservation of the adiabatic invariant
   tend to decrease the pitch angles, whereas the diffusion, that
   pushes back on electrons by means of the aforementioned waves,
   tends to increase the pitch angles. By examining the
   quasi-stationary state, we investigate a regime in which these two
   processes are balanced and a non-vanishing value of pitch
   angles is created.}
   {}


   \maketitle
%


\section{Introduction}

One of the major problems related to active galactic nuclei is the
origin of the nonthermal high energy radiation. According to
standard approaches, the most commonly encountered radiation
mechanisms at a level sufficient for application to AGN  is the
synchrotron mechanism and the inverse Compton scattering
(\cite{blan}). Because of strong synchrotron losses, relativistic
electrons in general, quickly lose their perpendicular energy, and
on a synchrotron cooling timescale $\sim 10^{-6}s-10^{-3}s$, the
particles transit to their ground Landau state. In this case, the
electrons may be described approximately as moving one-dimensionally
along the field lines and the synchrotron radiation to have been
absorbed. This is why the broadband emission spectrum of AGN
consists of two components: the high-energy (from X-rays to
$\gamma$-rays) component is formed by the inverse Compton scattering
and not by the synchrotron mechanism, which is supposed to be
involved only in the low-energy (from radio to optical/UV) band.
However, under certain conditions, due to the QLD of cyclotron
waves, the pitch angles might increase, leading to the efficient
production of synchrotron radiation. 

The QLD was applied to pulsars in a series of papers
(\cite{machus1,lomin,malmach}). Malov \& Machabeli (2001) studied
optical synchrotron emission of radio pulsars. In the outer parts of
pulsar magnetospheres, these authors demonstrated that because of
the cyclotron instability, the transverse momenta of relativistic
particles is non-zero, giving rise to the pitch angle distribution,
which in turn, via the QLD, leads to the synchrotron emission.
Applying the kinetic approach to a particular pulsar, RX
J1856.5-3754, \cite{nino} showed that waves excited by the cyclotron
mechanism, in terms of the creation of the pitch angles, come into
the radio domain. The QLD interesting because the recent detection
of very high energy (VHE) pulsed emission form the Crab pulsar
(\cite{magic}). The MAGIC Cherenkov telescope discovered the pulsed
emission above 25GeV between 2007 October and 2008 February. It has
been shown that the corresponding VHE signal peaks at the same phase
as the signal in the optical spectrum (\cite{magic}). In turn this
indicates that the polar cap models must be excluded from the
possible scenario of the radiation. On the other hand, analysis of
the MAGIC data implies that the location of the aforementioned VHE
and optical radiation must be the same. According to the
quasi-linear diffusion, on length scales typical of the light
cylinder (a hypothetical zone, where the linear velocity of rigid
rotation equals exactly the speed of light), the cyclotron
instability occurs in the optical band, leading to an increase in
the pitch angles via the QLD. This mechanism automatically explains
the coincidence of phases in the optical and VHE bands
(\cite{difus}).

AGN magnetospheres are supported by strong magnetic fields and
therefore, the QLD might also be of great importance to these
particular objects. As aforementioned, for ultra-relativistic
electrons the synchrotron losses are so efficient that the
synchrotron mechanism takes place only for relatively low energy
particles and highly relativistic electrons are involved in
radiation via the inverse Compton scattering. This is not true for
the QLD, because as for the pulsar magnetospheres, AGN
magnetospheric particles will undergo the QLD, preventing the rapid
damping of pitch angles, giving rise to the emission process.

In the present paper, we study the role of the QLD in producing the
X-rays via the synchrotron mechanism in AGN magnetospheres. The
paper is organized as follows. In Section 2, we consider the kinetic
approach to the quasi-linear diffusion, in Sect. 3 we present our
results and in Sect. 4 we summarize them.

\section{Main consideration} \label{sec:consid}

When relativistic particles move in the magnetic field, they emit
electromagnetic waves corresponding to the photon energies
(\cite{Lightman})
\begin{equation}
\label{eps} \epsilon_{keV}\approx 1.2\times
10^{-11}B\gamma^2\sin\psi,
\end{equation}
where by $B$ we denote the magnetic induction, $\gamma$ is the
Lorentz factor of particles, and $\psi$ denotes the pitch angle.
Equation (\ref{eps}) represents the photon energy in the maximum
emission intensity. As we already mentioned in the introduction, the
timescale is very short for the transit to the ground Landau state
that provides quasi-one-dimensional motion of electrons along the
field lines without radiation. As investigated in a series of papers
(\cite{machus1,malmach}), the cyclotron instability of the
electron-positron plasma under certain conditions,  may however
"create" pitch angles, which activate the subsequent synchrotron
process.

We consider the plasma to consist of two components: (a) the
electron-positron plasma component with the Lorentz factor,
$\gamma_p$ and (b) highly relativistic electrons, the so-called beam
component with the Lorentz factor, $\gamma_b$
($\gamma_b\gg\gamma_p$). According to the the QLD model, the
consequent transverse modes generate frequencies (\cite{kmm})
\begin{equation}\label{disp1}
\omega_t \approx kc\left(1-\delta\right),
\end{equation}
where
\begin{equation}\label{delta}
\delta = \frac{\omega_p^2}{4\omega_B^2\gamma_p^3}.
\end{equation}
We denote by $k$ the modulus of the wave vector, where $c$ is the
speed of light, $\omega_p \equiv \sqrt{4\pi n_pe^2/m}$ is the plasma
frequency, $\omega_B\equiv eB/mc$ is the cyclotron frequency, $e$
and $m$ are the electron charge and the rest mass, respectively, and
$n_p$ is the plasma density.


Kazbegi et al. (1992) demonstrated that the aforementioned waves are
excited if the cyclotron resonance condition
\begin{equation}\label{cycl}
\omega - k_{_{\|}}V_{_{\|}}-k_xu_x\pm\frac{\omega_B}{\gamma_b} = 0,
\end{equation}
is satisfied, where $u_x\equiv cV_{_{_{\|}}}\gamma_b/\rho\omega_B$
denotes the drift velocity of particles, $k_{_{\|}}$ is the wave
vector's longitudinal (parallel to the background magnetic field)
component, $k_x$ is the wave vector's component along the drift,
$V_{_{\|}}$ is the component of velocity along the magnetic field
lines, and $\rho$ is field line's curvature radius. By taking into
account the resonance condition, from Eq. (\ref{disp1}), one can
obtain an expression for the excited cyclotron frequency
\begin{equation}\label{om1}
\omega\approx \frac{\omega_B}{\delta\gamma_b}.
\end{equation}
In deriving Eq. (\ref{om1}), we have taken into account the
condition $\lambda>n_p^{-1/3}$, which means that in exciting waves
all resonance particles participate (collective phenomena),
therefore the range of spectral frequencies is wide. In general, the
cyclotron mode excites if the distribution function, $f$,  is almost
one dimensional and depends on the longitudinal momentum. On the
other hand, the magnetic field in AGN magnetospheres is strong
enough to maintain the frozen-in condition, which in turn means that
the particles follow field lines and thus, $f$ behaves according to
$p_{||}$.

When particles move in a nonuniform magnetic field, they undergo a
force ${\bf G}$ that is responsible for the conservation of the
adiabatic invariant, $I = 3cp_{\perp}^2/2eB$ (\cite{landau}). The
corresponding components of this force are given by
\begin{equation}\label{g}
G_{\perp} = -\frac{mc^2}{\rho}\gamma_b\psi,\;\;\;\;\;G_{_{\|}} =
\frac{mc^2}{\rho}\gamma_b\psi^2.
\end{equation}
In the synchrotron regime, we should detect the radiative force
(\cite{landau}):
\begin{equation}\label{f}
F_{\perp} = -\alpha\psi(1 + \gamma_b^2\psi^2),\;\;\;\;\;F_{_{\|}} =
-\alpha\gamma_b^2\psi^2,
\end{equation}
where $\alpha = 2e^2\omega_B^2/(3c^2)$.

These forces (${\bf F}$, ${\bf G}$) tend to decrease the pitch angle
of the particle. On the other hand, the feedback of low frequency
waves excited by particles be means of the cyclotron resonance,
leads to the quasi-linear diffusion of particles.  In turn, the QLD,
attempts to widen the range of the pitch angles opposing both ${\bf
F}$ and ${\bf G}$. The dynamical process saturates when the effects
of the above-mentioned forces are balanced by the diffusion. There
are, in general two different mechanisms of radiation: (I) the
resonance cyclotron emission and (II) the synchrotron process, the
first of which, as we have already mentioned, is a collective
phenomenon ($\lambda>n_p^{-1/3}$), whereas the second is a single
particle process ($\lambda<n_p^{-1/3}$) that does not require
superposition.

We consider the case $|G_{\perp}|\gg |F_{\perp}|$ and
$|G_{_\parallel}|\ll |F_{_\parallel}|$. By assuming a
quasi-stationary scenario ($\partial/\partial t = 0$), the
corresponding kinetic equation can be given by (\cite{malmach})
$$\frac{1}{mc\gamma_b\psi}\frac{\partial}{\partial\psi}
\left(\psi G_{_\perp}f\right) +
\frac{1}{mc}\frac{\partial}{\partial\gamma_b}\left(F_{_\parallel}
f\right) + \upsilon\frac{\partial f}{\partial r}=$$

$$=\frac{1}{m^2c^2\gamma_b^2\psi}\frac{\partial}{\partial\psi}
\left(\psi D_{_{\perp\perp}}\frac{\partial
f}{\partial\psi}\right)+\frac{1}{mc\psi}\frac{\partial}{\partial\psi}
\left(\psi^2 D_{_{\perp\parallel}}\frac{\partial
f}{\partial\gamma_b}\right)+$$

\begin{equation}\label{kinet}
\;\;\;\;\;\;\;\;+\frac{1}{mc}\frac{\partial}{\partial\gamma_b}
\left(\psi D_{_{\perp\parallel}}\frac{\partial
f}{\partial\psi}\right),
\end{equation}
where $f = f(\psi,p{_{_\parallel}})$ is the distribution function of
particles, $p_{_\parallel}$ is the longitudinal momentum,
\begin{equation}\label{dif}
D_{\perp\perp}\approx \frac{\pi
e^2}{4c}\frac{\omega_p^2}{\omega_B^2}\gamma_b|E_k|^2,\;\;\;\;\;\;
D_{\perp_{\parallel}}\approx -\frac{\pi e^2}{4cp_\parallel}|E_k|^2,
\end{equation}
are the diffusion coefficients, $n_b$ is the density of the beam
component, and $|E_k|^2$ is the energy density per unit wavelength.
The corresponding energy density can be estimated to be $|E_k|^2k$.
We assume that half of the plasma energy density,
$mc^2n_b\gamma_b/2$ is converted into the waves, such that $|E_k|^2$
is given by
\begin{equation}\label{ek2}
|E_k|^2 = \frac{mc^3n_b\gamma_b} {2\omega}.
\end{equation}
By expressing the distribution function as
$\chi(\psi)f(p_{_\parallel})$, one can solve Eq. (\ref{dif}) for
$\chi$

\begin{equation}\label{chi} \chi(\psi) = C_1e^{-A\psi^2},
\end{equation}
where
\begin{equation}\label{A}
A\equiv \frac{2m^2c^4\gamma_b^2\left(\omega_B/\omega_p\right)^2}
{\pi e^2\rho|E_k|^2\gamma_p},
\end{equation}

The corresponding mean value of the pitch angle can be estimated as
to be
\begin{equation}\label{pitch}
\bar{\psi}
 = \frac{\int_{0}^{\infty}\psi\chi(\psi)d\psi}{\int_{0}^{\infty}\chi(\psi)d\psi}
= \frac{1}{\sqrt{\pi A}}.
\end{equation}
Therefore, as we see, the QLD maintains the pitch angles and
prevents them from damping, which in turn maintains the properties
of the synchrotron process.

\section{Discussion}
\begin{figure}
  \resizebox{\hsize}{!}{\includegraphics[angle=0]{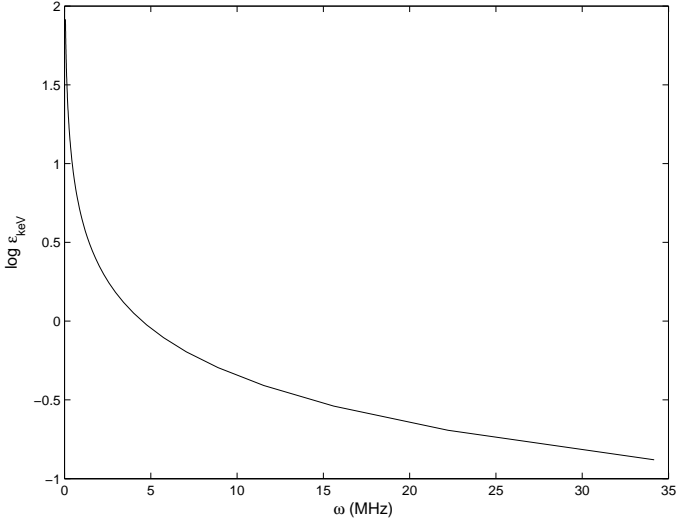}}
  \caption{The synchrotron emission energy versus the cyclotron
  frequency. The set of parameters is $M_{BH} =
10^8M_{\odot}$, $n_b = 2000cm^{-3}$, $B = 10^4G$, $\gamma_p = 2$,
$\gamma_b\in\{4\times 10^7;10^9\}$, and $\rho = R_g$. As we see from
the plot, the low frequency cyclotron mechanism arises in the radio
band, inducing the QLD and subsequently creating relatively high
pitch angles, which in turn lead to the X-ray emission.}\label{fig1}
\end{figure}
Most AGN exhibit VHE emission, which in turn indicates that the AGN
magnetosphere is contained of ultra-relativistic electrons. In this
context, the origin of acceleration of particles is very important.
Proposed mechanisms such as Fermi-type acceleration (\cite{cw99}),
centrifugal acceleration (\cite{mr94,osm7,ra08}), and acceleration
due to the black hole dynamo mechanism (\cite{levins}) can
effectively provide Lorentz factors of the order of $\sim 10^{5-9}$.

For an isotropic distribution of relativistic electrons
(\cite{blan}) in very strong magnetic fields, particles emit in the
synchrotron regime with the power, $P\approx
2e^4B^2\beta_{\perp}^2\gamma^2/(3m^2c^3)$, therefore by assuming
$\beta_{\perp}\sim 1$, the synchrotron cooling timescale,
$t_s\equiv\gamma mc^2/P$ can be estimated to be
\begin{equation}\label{ts}
t_s\approx 5\times
10^{-6}\times\left(\frac{10^4G}{B}\right)^2\times\left(\frac{10^7}{\gamma}\right)s,
\end{equation}
where we have taken into account that the energy is uniformly
distributed between the beam and the plasma components,
$n_b\gamma_b\approx n_p\gamma_p$,
\begin{figure}
  \resizebox{\hsize}{!}{\includegraphics[angle=0]{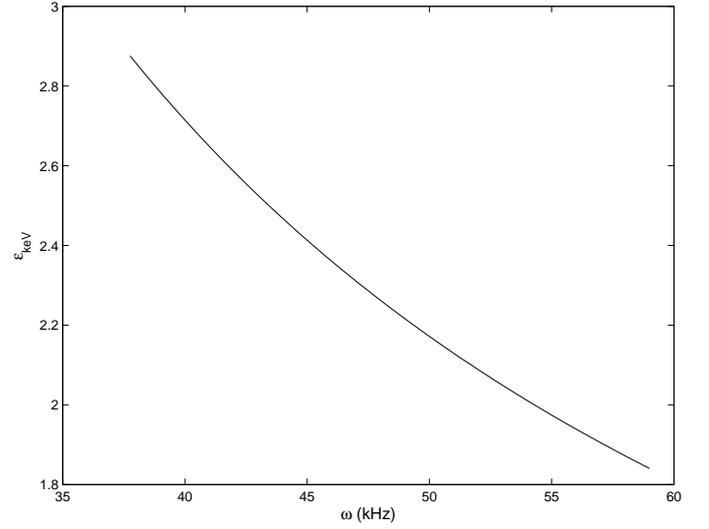}}
  \caption{The synchrotron emission energy versus the cyclotron
  frequency. The set of parameters is $n_b = 2000cm^{-3}$, $L = 10^{45}erg/s$, $\Omega =
3\times 10^{-5}s^{-1}$, $\gamma_p = 2$, $\gamma_b\in\{4;5\}\times
10^6$ and $\rho = r_{lc}$. As is clear, the low frequency cyclotron
mechanism arises in the radio band, inducing via the QLD, the X-ray
emission.}\label{fig2}
\end{figure}
and the magnetic field is normalized to the typical value of the
magnetic induction close to the supermassive black hole
(\cite{paradig}). As is clear from this expression, the synchrotron
timescale for ultra-relativistic electrons in the very vicinity of
the black hole is so short ($\sim 10^{-5}s$) that these particles
very rapidly lose their perpendicular kinetic energy and transit to
the ground Landau state. One can obtain a similar result for outer
magnetospheric lengthscales, thus for the light cylinder area, when
the magnetic field is determined by the bolometric luminosity, $L$
of AGN to be the equipartition field
\begin{equation}
\label{b} B_{lc}\approx \sqrt{\frac{2L}{r_{lc}^2c}}\approx
290\times\left(\frac{L}{10^{45}erg/s}\right)^{1/2}\times\left(\frac{\Omega}{3\times
10^{-5}s^{-1}}\right)G,
\end{equation}
where $r_{lc} = c/\Omega$ is the light cylinder radius and $\Omega$
the angular velocity of rotation. If we consider a typical value of
the bolometric luminosity, $\sim 10^{45}erg/s$, then one can see
from Eq. (\ref{b}) that the magnetic induction equals approximately
$290G$, which, combined with Eq. (\ref{ts}), leads to the cooling
timescale, $5.6\times 10^{-4}s$.

All quantities vary because of the cyclotron instability, which
causes to the QLD. By using Eqs. (\ref{delta}-\ref{om1}), we can
estimate the frequency of the cyclotron mode to be
\begin{equation}
\label{om} \omega\approx 6.2\times
10^8\left(\frac{\gamma_p}{2}\right)^4\left(\frac{10^7}
{\gamma_b}\right)^2\left(\frac{B}{10^4G}\right)^3\left(\frac{2\times
10^3cm^{-3}}{n_b}\right)Hz,
\end{equation}
We consider a nearby zone of the AGN with mass $M_{BH} =
10^9M_{\odot}$ ($M_{\odot}$ is the solar mass) and typical
magnetospheric parameters $n_b = 2000cm^{-3}$, $B = 10^4G$ and
$\gamma_p = 2$. Then, examining the ultra-relativistic beam
component electrons with Lorentz factors $\gamma_b\in\{3;4\}\times
10^7$ and assuming that the curvature radius $\rho$ of field lines
is of the order of the gravitational radius, $R_g\equiv
2GM_{BH}/c^2$, one can see from Eq. (\ref{pitch}) that the  created
pitch angle is of the order of $10^{-9}rad$, which indicates the
synchrotron emission in the $X$-ray domain [see Eq. (\ref{eps})]. In
Fig. \ref{fig1}, we show the synchrotron emission energy versus the
excited cyclotron frequency. As is clear from the figure, the
cyclotron instability appears in the radio band leading to the
$X$-ray emission. One can see that the radio emission is generated
by the collective phenomena. We consider the radio frequency $5MeV$,
corresponding to the wavelength, $\lambda$, of the order of $6000
cm$. On the other hand, the average distance between particles,
$d=n_p^{-1/3}$ is of the order of $2\times 10^{-4}cm$, which is
shorter by many orders of magnitude than $\lambda$. This in turn,
indicates that emission in the radio band is provided by the
collective phenomena. In contrast to this, the VHE radiation
generated in the $X$-ray domain (see Fig. \ref{fig1}) is a single
particle mechanism. The importance of the quasi-linear diffusion is
twofold: (a) it generates the synchrotron radiation in strong
magnetic fields for ultra-relativistic particles, which would be
impossible without the QLD and (b) it simultaneously excites
emission in two different domains. In this context, it is easy to
check whether the emission is driven by the QLD by verifying that
(I) the radiation is linearly polarized and (II) both signals have
equal phases. In (\cite{difus}), we performed the same theoretical
analysis of the observed VHE ($>25GeV$) pulsed emission of the Crab
pulsar (\cite{magic}).

Unlike the previous case, we show in Fig. \ref{fig2} the same
behavior for the outer magnetospheric (light cylinder) lengthscales.
The set of parameters is $M_{BH} = 10^9M_{\odot}$, $n_b =
2000cm^{-3}$, $L = 10^{45}erg/s$, $\Omega = 3\times 10^{-5}s^{-1}$,
$\gamma_p = 3$, $\gamma_b\in\{3;4\}\times 10^7$ and $\rho = r_{lc}$.
As shown in Fig. \ref{fig2}, the radio frequency close to the light
cylinder zone, in the kHz domain excites the hard $X$-ray emission
by means of the QLD. From Eqs. (\ref{g},\ref{f}), one can
straightforwardly check the validity of our assumptions,
$|G_{\perp}|\gg |F_{\perp}|$ and $|G_{_\parallel}|\ll
|F_{_\parallel}|$, confirming our approach.
\begin{figure}
  \resizebox{\hsize}{!}{\includegraphics[angle=0]{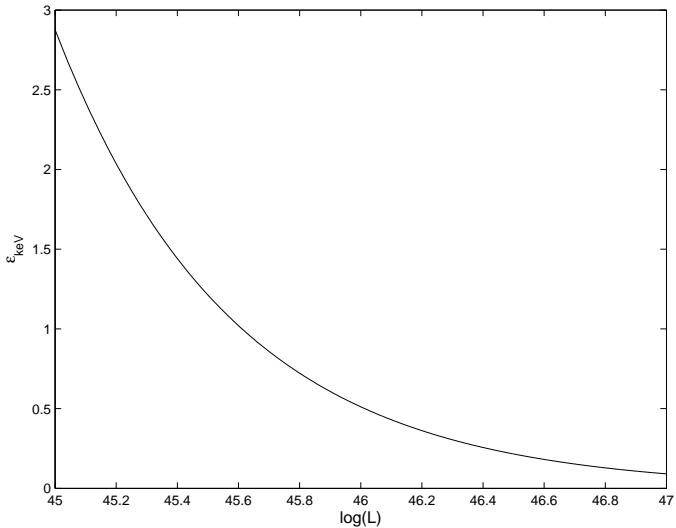}}
  \caption{The synchrotron emission energy versus the AGN luminosity.
  The set of parameters is $n_b = 2000cm^{-3}$, $L = 10^{45-47}erg/s$, $\Omega
= 3\times 10^{-5}s^{-1}$, $\gamma_p = 2$, $\gamma_b = 5\times 10^6$
and $\rho = r_{lc}$. }\label{fig3}
\end{figure}
It is also interesting to investigate the QLD for different values
of the luminosity. We examine the luminosity interval from
$10^{45}erg/s$ to the Eddington limit, which for the given black
hole mass, $10^9M_{\odot}$, equals $10^{47}erg/s$. In Fig.
\ref{fig3}, we show the strong $X$-ray energy versus the AGN
luminosity. The set of parameters is $M_{BH} = 10^9M_{\odot}$, $n_b
= 2000cm^{-3}$, $L = 10^{45-47}erg/s$, $\Omega = 3\times
10^{-5}s^{-1}$, $\gamma_p = 3$, $\gamma_b = 3\times 10^7$ and $\rho
= r_{lc}$. As is evident from the figure, the emission energy is a
continuously decreasing function of the luminosity. Indeed, by
combining Eqs. (\ref{eps},\ref{A},\ref{pitch},\ref{b}), we see that
the synchrotron emission energy behaves as $L^{-3/4}$.

Therefore, as our investigation shows, the QLD is a working
mechanism in AGN magnetospheric flows and drives the synchrotron
process.

\section{Summary}\label{sec:summary}

The main aspects of the present work can be summarized as follows:
\begin{enumerate}

      \item We studied the quasi-linear interaction of proper modes of
      AGN magnetospheric plasmas with the resonant plasma particles.
      For this purpose, the synchrotron reaction
      force has been taken into account. The role of the QLD was studied in
      the context of producing the soft and hard $X$-ray emission from AGN.

      \item It has been shown that the synchrotron cooling
      timescales for ultra-relativistic electrons are very small,
      and particles rapidly transit to the ground
      Landau state, that in turn, prevents the subsequent radiation.
      We found that, under certain conditions the cyclotron
      instability develops, leading to the creation of pitch angles,
      and the subsequent synchrotron process.

      \item We have considered two extreme regions of magnetospheres:
      (a) relatively close to the black hole and
      (b) the light cylinder zone.
      As our model shows, the cyclotron instability, under certain conditions,
      generates the radio frequency in the range $(0.06-35)MHz$
      and creates the soft $X$-ray emission, $(0.13-100)keV$, via the QLD.
      For the light cylinder area, radio spectra occurs in the range
      $(38-59)kHz$, which produces the hard $X$-ray emission
      in the domain $(1.8-2.9)keV$. We have emphasized that
      from an observational evidence one can directly verify the
      validity of the QLD by determining (I) the polarization and (II)
      phases of signals in radio and $X$-ray domains respectively.

      \item The quasi-linear diffusion has also been studied versus
      the AGN luminosity. It was shown that for more luminous AGN,
      the corresponding photon energy of the hard $X$-ray emission,
      generated by the synchrotron mechanism is lower.

      \end{enumerate}

\section*{Acknowledgments}
The research was supported by the Georgian National Science
Foundation grant GNSF/ST07/4-193.


\begin{thebibliography}{}


\bibitem[Albert et al. 2008]{magic} Abert J. et al., 2008, \apj, 674, 1037A
\bibitem[Blandford et al. 1990]{blan} Blandford R.D., Netzer H. \&
Woltjer L., 1990, Active Galactic Nuclei, Springer-Verlag
\bibitem[Catanese \& Weeks 1999]{cw99} Catanese M. \& Weeks T.C. 1999, PASP, 111, 1193
\bibitem[Chkheidze 2009]{nino1} Chkheidze N., 2009, A\&A, 500,
861
\bibitem[Chkheidze \& Lomiashvili 2008]{nino2} Chkheidze N. \& Lomiashvili
G., New. Ar., 2008, 13, 12
\bibitem[Chkheidze \& Machabeli (2007)]{nino}
Chkheidze N. \& Machabeli G., 2007, A\&A, 471, 599
\bibitem[Kazbegi et al. 1992]{kmm} Kazbegi A.Z., Machabeli G.Z
\& Melikidze G.I., 1992, \mnras, 253, 377
\bibitem[Landau \& Lifshitz 1971]{landau} Landau L.D. \& Lifshitz E.M.,
51971, Classical Theory of Fields (London: Pergamon)
\bibitem[Levinson 2000]{levins} Levinson Amir, 2000, Phys.Rev.L, 85, 912
\bibitem[Lominadze et al. 1979 ]{lomin} Lominadze J.G., Machabeli G.Z.
\& Mikhailovsky A.B., 1979, J. Phys. Colloq., 40, No. C-7, 713
\bibitem[Machabeli \& Osmanov 2009 ]{difus} Machabeli G. \& Osmanov
Z., 2009, \apj L, 700, 114
\bibitem[Machabeli \& Rogava 1994]{mr94} Machabeli G.Z. \& Rogava A.D., 1994, Phys. Rev. A, 50, 98
\bibitem[Machabeli \& Usov 1979]{machus1} Machabeli G.Z. \& Usov
V.V., 1979, AZhh Pis'ma, 5, 238
\bibitem[Malov \& Machabeli 2001]{malmach} Malov I.F. \&
Machabeli G.Z., 2001, \apj, 554, 587
\bibitem[Osmanov et al. 2007]{osm7} Osmanov Z., Rogava A.S. \& Bodo G., 2007, A\&A, 470, 395
\bibitem[Rieger \& Aharonian 2008]{ra08} Rieger F.M., \& Aharonian F.A., 2008, A\&A,
479, L5
\bibitem[Rybicki \& Lightman 1979]{Lightman} Rybicki  G.B. \& Lightman A. P., 1979,
Radiative Processes in Astrophysics. Wiley, New York
\bibitem[Thorne et al. 1988]{paradig} Thorne K.S., Price R.H. \& Macdonald D.A., 1988,
Black Holes: The Membrane Paradigm, Yale University Press, New
Haven.


\end{thebibliography}
\end{document}